\documentclass[fleqn,usenatbib]{mnras}

\usepackage{newtxtext,newtxmath}

\usepackage[T1]{fontenc}
\usepackage{ae,aecompl}

\makeatletter
\newcommand{\@todonotes@enable}{1}
\newcommand{\@todonotes@inline}{1}
\makeatother
\input{notes.tex}

\usepackage{graphicx}	% Including figure files
\usepackage{amsmath}	% Advanced maths commands
\usepackage{amssymb}	% Extra maths symbols

\newcommand\alphafit{-2.7 }
\newcommand\alphafiterror{0.4 }
\newcommand\alphafitevolve{-3.1 }
\newcommand\alphafitevolveerror{0.6 }

\newcommand\rhlogfit{0.44 }
\newcommand\rhlogfiterror{0.04 }
\newcommand\rhfit{2.7 }
\newcommand\rhfiterror{0.3 }

\newcommand\rhfitevolve{3.26 }

\newcommand\sigfit{1.7 }
\newcommand\sigfiterror{0.2 }

\newcommand\rmlogfit{0.41 }
\newcommand\rmlogfiterror{0.01 }
\newcommand\rmsiglogfit{0.20 }
\newcommand\rmsiglogfiterror{0.01 }

\newcommand\rmfit{2.57 }
\newcommand\rmfiterror{0.04 }
\newcommand\rmsigfit{1.59 }
\newcommand\rmsigfiterror{0.01 }

\title[M83]{The Initial Properties of Young Star Clusters in M83}

\author[Webb \& Sills]{
Jeremy J. Webb$^{1}$\thanks{E-mail: jeremy.webb@utoronto.ca}
\& Alison Sills$^{2}$
\\
$^{1}$Department of Astronomy and Astrophysics, University of Toronto, 50 St. George Street, Toronto, ON, M5S 3H4, Canada\\
$^{2}$Department of Physics \& Astronomy, McMaster University, 1280 Main Street West, Hamilton, ON, L8S 4M1, Canada\\
}

\date{Accepted XXX. Received YYY; in original form ZZZ}

\pubyear{2020}

\begin{document}
\label{firstpage}
\pagerange{\pageref{firstpage}--\pageref{lastpage}}
\maketitle

% Abstract of the paper (244 words)
\begin{abstract}
The initial sizes and masses of massive star clusters provide information about the cluster formation process and also determine how cluster populations are modified and destroyed, which have implications for using clusters as tracers of galaxy assembly. Young massive cluster populations are often assumed to be unchanged since cluster formation, and therefore their distribution of masses and radii are used as the initial values. However, the first few hundred million years of cluster evolution does change both cluster mass and cluster radius, through both internal and external processes. In this paper, we use a large suite of $N$-body cluster simulations in an appropriate tidal field to determine the best initial mass and initial size distributions of young clusters in the nearby galaxy M83. We find that the initial masses follow a power-law distribution with a slope of \alphafit $\pm$ \alphafiterror, and the half-mass radii follow a log-normal distribution with a mean of \rmfit\ $\pm$ \rmfiterror\ pc and a dispersion of \rmsigfit\ $\pm$ \rmsigfiterror\ pc. The corresponding initial projected half-light radius function has a mean of \rhfit\ $\pm$ \rhfiterror\ pc and a dispersion of \sigfit\ $\pm$ \sigfiterror\ pc.  The evolution of the initial mass and size distribution functions are consistent with mass loss and expansion due to stellar evolution, independent of the external tidal field and the cluster's initial density profile. Observed cluster sizes and masses should not be used as the initial values, even when clusters are only a few hundred million years old.
\end{abstract}


%%%%%%%%%%%%%%%%%%%%%%%%%%%%%%%%%%%%%%%%%%%%%%%%%%

%%%%%%%%%%%%%%%%% BODY OF PAPER %%%%%%%%%%%%%%%%%%

\section{Introduction}

Star clusters are formed in the gas-rich regions of galaxies, and subsequently found in many galactic environments. Star clusters are capable of both shaping, and tracing, galaxy evolution. They are often used as tracers of the assembly history of the galaxy \citep[e.g.][]{Kruijssen2019, Massari2019}, and are also used to probe the environment in which they form \citep{Grudic2020}. For example, observations indicate that the cluster formation efficiency, defined as the fraction of stars which remain in bound clusters, depends on whether the clusters are located in or between spiral arms \citep[e.g.][]{Messa2018}, and that the age distribution of clusters depends on the density and velocity dispersion of the local interstellar medium \citep{Miholics2017}. The dissolution of clusters may be an important contribution to building up the halo and bulge population of galaxies \citep[e.g.][]{Martell2011, Perets2014}. Star clusters have also been proposed as a significant source of ionization radiation in the early universe \citep[e.g.][]{BK2018}, and the clustering of massive stars and their subsequent supernovae play an important role in regulating star formation in a galaxy \citep[e.g.][]{Gentry2017}. 

There are still many open questions regarding the earliest evolution of star clusters. Most of the massive clusters in the Milky Way are the old globular clusters, whose birthplaces are too far away in both space and time for us to investigate now. The youngest clusters are usually low mass, and are still deeply embedded in their natal clouds, making it difficult to clearly observe their properties. There are a few local galaxies with large populations of young massive clusters, such as those probed by the LEGUS survey \citep{Calzetti2015}, for which it is possible to determine fundamental properties such as mass, age, and size of the clusters, as well as their location within the galaxy. These observations provide the best samples to determine the initial properties of cluster populations. 

%\newpage

To use clusters as tracers of galaxy formation and evolution, we need to understand how clusters evolve with time. Massive clusters are subject to internal dynamical evolution caused by two-body relaxation, coupled with the effects of a tidal field imposed by the galaxy \citep[e.g.][]{Chernoff1990}. They can be subjected to tidal shocks as they orbit through a galactic disk \citep{Gnedin1997}, or if they encounter a massive giant molecular cloud \citep{Gieles06}. These processes, in addition to mass loss from individual stars, cause the clusters to lose mass and, under some circumstances, dissolve entirely. In order to properly determine the total effect of all these processes on the cluster population of a galaxy, we need to know the initial distributions of cluster properties, particularly the initial cluster mass and radius.  

The current state of our understanding of properties of young clusters have been reviewed recently by \citet{Adamo2020} and \citet{Krumholz2019}. We will briefly summarize the relevant points here, and refer the reader to those reviews for more details. The initial mass function of young clusters can be described by a power law with a slope of about -2. There is increasing evidence that a high-mass cutoff is also required (also known as a Schechter function), and that cutoff mass may depend on galactic environment. Both the necessity of a cutoff mass and its value are debated in the literature. The difference between a pure power law and one with an exponential cutoff shows up only at the high mass end, where in either case there are only a few expected objects. The stochastic nature of star cluster formation only exacerbates this problem. 

The initial size distribution of clusters is more difficult to determine observationally. Most studies on individual galaxies conclude that cluster radii are (nearly) independent of cluster mass, with a mean value of about 3 pc. This value is consistent with the half-mass radii of Milky Way globular clusters, suggesting that the radius is also independent of age, although observations might suggest an increase in radius in the first 10-100 Myr of cluster evolution. Both gas expulsion and mass loss due to stellar evolution have been invoked to drive that early expansion.  

In this paper, we provide an alternative approach for determining the initial size and mass distributions for a population of star clusters that accounts for star cluster evolution. Here, we explicitly define ``initial" to mean the time after the cluster has cleared away its natal gas and is a single, spherical cluster -- this is likely about 5-10 Myr after the first star started to form \citep{Howard2017}. Our approach is to then dynamically ``rewind" an observed population of clusters by taking into account mass loss due to stellar evolution, internal two-body relaxation processes, and the external tidal influence of the galaxy on the clusters. We account for these processes by constructing detailed $N$-body models of young massive clusters in the tidal environment of the nearby spiral galaxy M83. We then present a method to compare those models to the present-day population of clusters in that galaxy to extract the true initial properties of the cluster population, particularly the distribution of masses and radii.

We have chosen to use M83 as our test case for this method because of the quality of the data on the young cluster population that exists in the literature, and because we have a good understanding of the galactic environment as well. We will discuss the observations and the $N$-body models in Section 2, as well as describing our method to compare the two. Our results are given in Section 3, and we discuss their implications in Section 4.

\section{Methods}

In order to estimate the initial mass and size distributions of the young massive cluster population of M83, we will directly compare the present day observed properties of the system to a large suite of direct $N$-body star cluster simulations. Using a Monte Carlo-based approach, we can use the simulations to determine the set of initial distributions that best reproduce the current mass, size, and age distributions of clusters in M83. In the following subsections, we describe in detail the observed M83 dataset, the suite of simulations with which the observations will be compared to, and the ABC-MCMC method that we will employ to determine the cluster properties at formation.

\subsection{Data}

The measured properties of observed young massive clusters in M83 are taken from \citet{Ryon2015}, who analyzed archive HST/WFC data from Program 11360 (PI: O'Connell) and Program 12513 (PI: Blair). The authors use GALFIT (Peng et al. 2002, 2010) to fit the surface brightness profiles of individual clusters with Elson, Fall \& Freeman \citep[EFF,][]{Elson1987} profiles. From the fit, \citet{Ryon2015} report the best fit power-law slope of the EFF profile $\eta$ and the effective (i.e. projected half-light) radius $r_h$. They also use the spectral energy distribution fitting technique of \citet{Adamo2010} to estimate the age and mass of each young massive cluster. 

We restrict our analysis to clusters with high quality measurements of their age and mass. More specifically, we only consider clusters with ages between 3 and 300 Myr, as age estimates of clusters older than 300 Myr have high uncertainties. Similarly, we only consider clusters with masses between $10^4$ - $5.5 \times 10^4 M_{\odot}$ as lower mass clusters are affected by incomplete sampling of the stellar initial mass function. Finally we require the clusters have density profiles with $\eta > 1.1$, as \citet{Ryon2015} notes that measured structural parameters with $\eta < 1.1$ are typically unreliable.

The final dataset consists of 101 observed young massive clusters with projected galactocentric distances between 0.5 and 6.3 kpc, with core and effective radii spanning 0.15 -- 5 pc and 0.2 -- 10 pc respectively. The mean effective radius of the observed distribution is 3.52 pc with a standard deviation of 2 pc.
The mass function of the observed cluster population has a power-law slope of $-2.45 \pm 0.6$, however the mass function does not resemble a power-law beyond 50 000 $M_{\odot}$. This break is consistent with \citet{Bastian2012}, who find that the outer mass function of M83 clusters has a truncation mass of 50 000 $M_{\odot}$. \citet{Bastian2012} also suggests the inner clusters have a truncation mass at 150 000 $M_{\odot}$. Our dataset does not probe such high masses. Below 50 000 $M_{\odot}$, the mass function has a power-law slope of $-3.15 \pm 0.6$. The density profiles of the clusters have EFF power-law indices between 1.1 - 9.0, with most having $\eta$ less than 3.0. The age distribution of the final list of clusters is flat between 3.0 and 265 Myr.

\subsection{N-body models}

The star cluster simulations were performed using the direct $N$-body code \texttt{NBODY6tt} \citep{renaud11}, a modified version of \texttt{NBODY6} \citep{aarseth03} that allows for arbitrary tidal fields to be used as the external galaxy potential. The tidal field of M83 is taken from \citet{Lundgren04}, who use the kinematics of molecular gas in M83 to estimate that the galaxy's potential is best represent by two exponential disks. The more massive disk has a mass of $5.9 \times 10^{10} M_{\odot}$ and a scale radius of 2.7 kpc, while the less massive disk has a mass of $3.0 \times 10^8 M_{\odot}$ and a scale radius of 0.05 kpc.

Masses of the stars within the clusters are generated assuming a \citet{Kroupa01} initial mass function between 0.1 and 50 M$_\odot$ and evolve in time according to the stellar evolution prescriptions of \citet{Hurley2000} for single stars and \citet{Hurley2002} for any binary stars that form. We initially assume each cluster has a metallicity of Z=0.001 and a primordial binary fraction of zero.  The positions and velocities of individual stars are generated assuming the cluster's density profile is either a Plummer model \citep{Plummer1911} or an EFF model \citep{Elson1987} and that the system is initially in virial equilibrium. We note that the clusters in M83 are metal-rich. As shown in \citet{Hurley2004}, the accompanying difference in stellar mass loss and therefore the dynamical effect is at most 1\%.

To generate a large suite of simulations that can be used to predict the initial masses and sizes of clusters in M83, it is essential to finely sample a large range of initial conditions. In order to estimate the range of initial masses and sizes that model clusters must have in order to reproduce the M83 cluster population, we first note that clusters will lose up to $30\%$ of their initial mass due to stellar evolution alone. Hence our model clusters must have initial masses of between 15,000 and 45,000 $M_{\odot}$ in order for their masses after 300 Myr of evolution to fall within the observed cluster range. This burst of mass loss also results in a reduction of the cluster's potential, which is known to cause clusters to undergo significant expansion early in their lifetime. Motivated by the stellar evolution-induced expansion observed in past simulations \citep[e.g.][]{Webb2014}, we estimate that clusters must have initial half-mass radii between 0.5 and 8 pc in order to reproduce the observed sizes of M83 clusters. 

The range of initial cluster orbits and density profiles that need to be considered are more difficult to constrain. It is not clear whether or not either property will play a role in the early evolution of young massive clusters. Therefore, we first investigate how strongly the first 300 Myr of cluster evolution depends on galactocentric distance and the initial density profile. To explore the effects of galactocentric distance on cluster evolution, we simulate clusters with circular orbits at  distances of 2 kpc, 4 kpc, and 6 kpc. We consider Plummer model clusters initially consisting of 25 000, 37 000, 50 000 and 75 000 stars and with half-mass radii of 6 pc. Only a larger cluster size is included as extended clusters are more strongly affected by the galactic tidal field than compact clusters. 

\begin{figure}
    \includegraphics[width=0.48\textwidth]{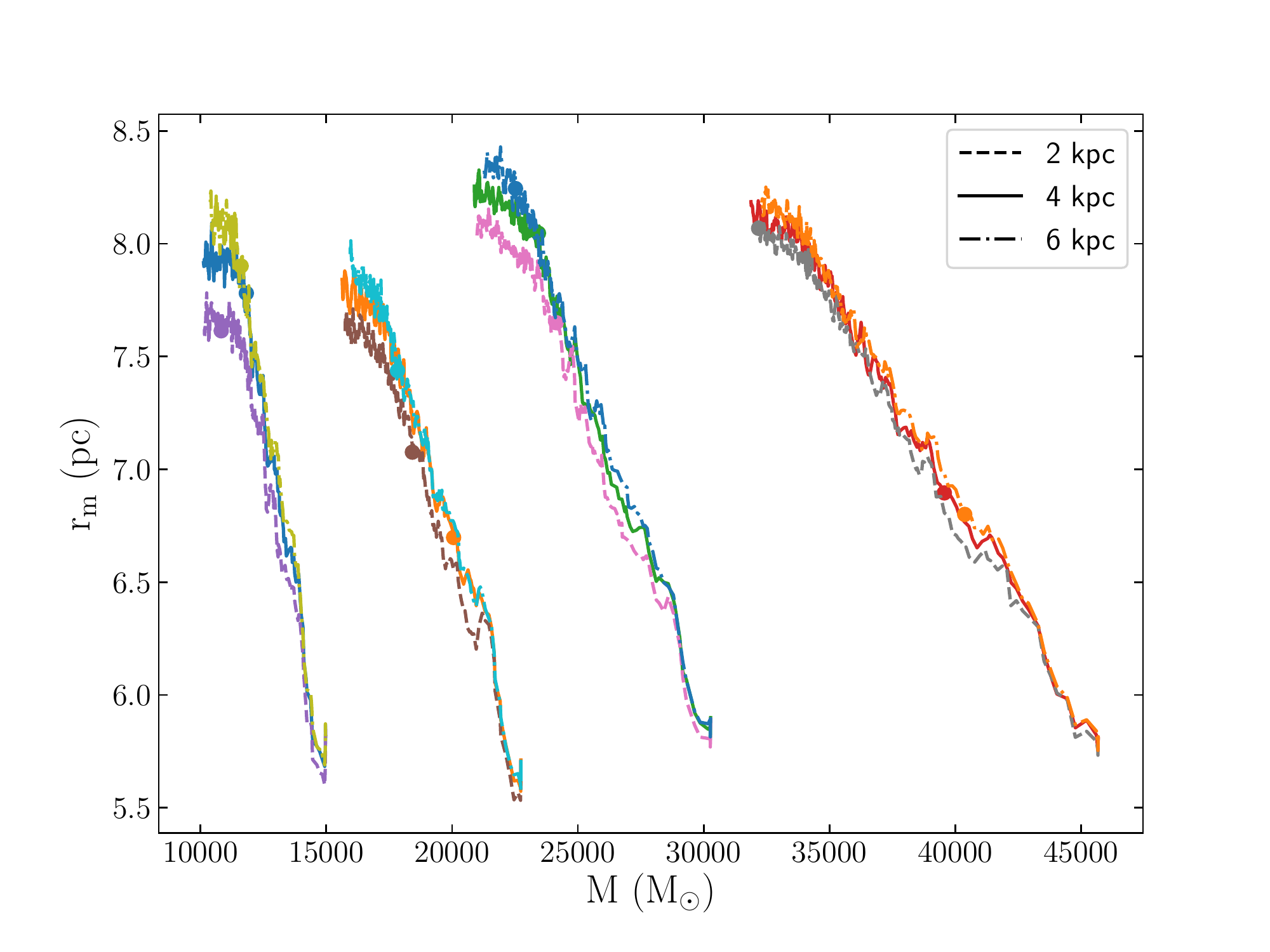}
    \caption{Mass and half-mass radius evolution of model clusters orbiting at 2 kpc (dashed line), 4 kpc (solid line) and 6 kpc (dashed-dot line). The mass and size of evolution of clusters in M83 shows no dependence on galactocentric distance within 300 Myr of birth.}
    \label{fig:rgcplot}
\end{figure}

Figure \ref{fig:rgcplot} illustrates the mass and size evolution of our test suite of extended cluster simulations with circular orbits at 2 kpc, 4 kpc, and 6 kpc. We find that the early evolution of young massive clusters is independent of galactocentric distance in M83 over all initial cluster masses. The ages of the clusters are less than one half-mass relaxation time, and therefore stars have not yet had time to escape.  Clusters with initial sizes less than 6 pc will also follow this trend as they will be even less tidally affected than the 6 pc clusters. Hence we can remove galactocentric distance from the list of parameters we need to consider in order to reproduce the observed clusters in M83 and just set all clusters to orbit at the intermediate distance of 4 kpc.

Similarly, we investigate the dependence of early cluster evolution on the cluster's initial density profile. We consider clusters initially consisting of 37,000 stars orbiting at a galactocentric distance of 4 kpc and with initial half-mass radii of 0.5 pc, 1.0, pc, 2.0 pc, 4.0 pc, and 6.0 pc. The initial density profiles of the test clusters are set to either a Plummer model or an EFF power-law model with $\eta$ equalling 1.1, 1.5, 2.0, or 2.5. In the EFF power-law model cases, the range of $\eta$ values encompasses the range observed in M83. 

\begin{figure}
    \includegraphics[width=0.48\textwidth]{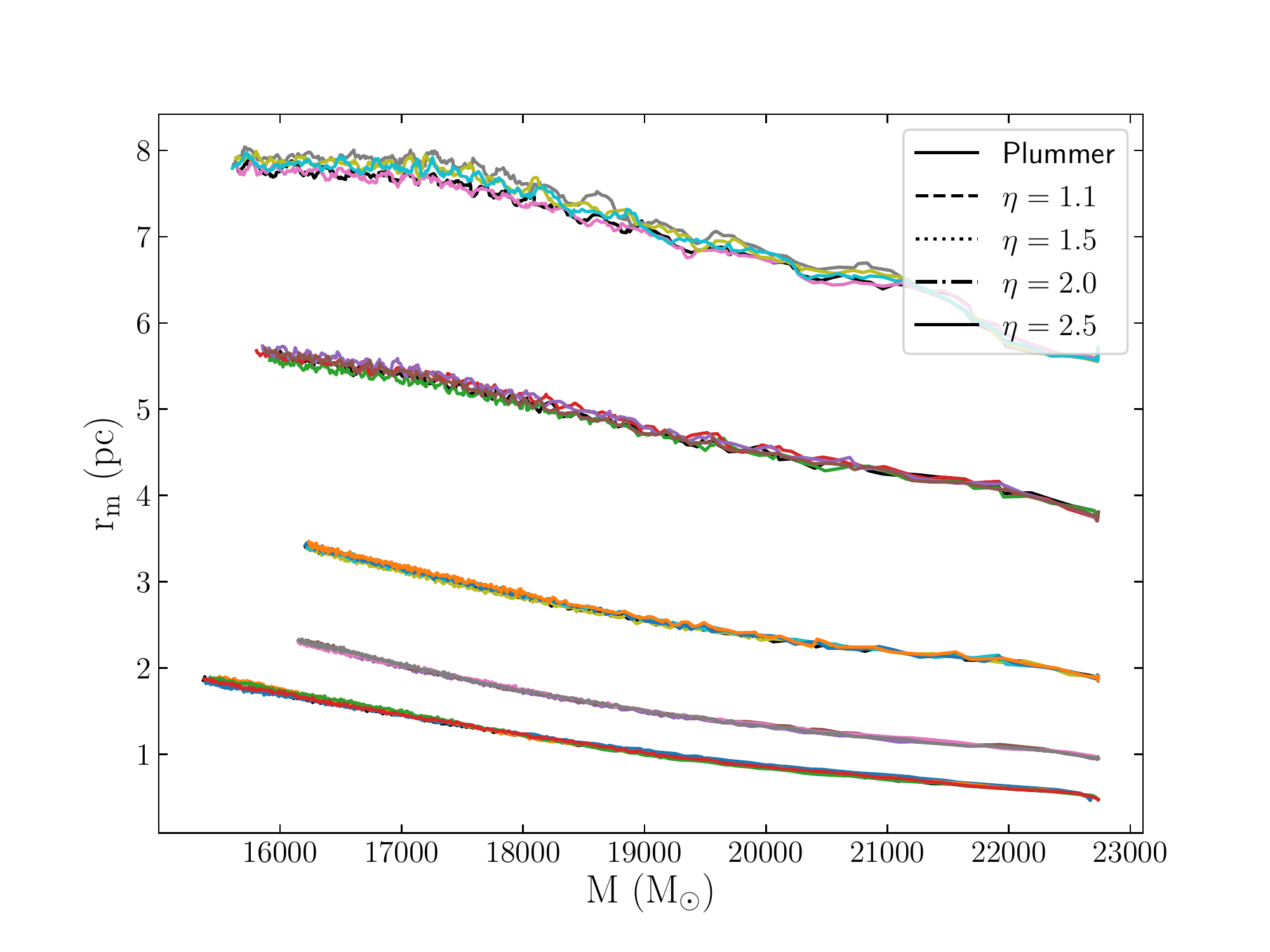}
    \caption{Mass and half-mass radius evolution of model clusters with different initial half-mass radii and density profiles orbiting at 4 kpc that initially consist of 37,000 stars. In addition to clusters initialized as Plummer spheres (solid line), we consider clusters initialized using EFF density profiles where the $\eta$ parameter is shown in the legend. The mass and size of evolution of clusters in M83 shows no dependence on their initial density profile within 300 Myr of birth.}
    \label{fig:etaplot}
\end{figure}

Figure \ref{fig:etaplot} illustrates the mass and size evolution of the second test suite of cluster simulations. The test models indicate that the early evolution of a cluster in M83's tidal field is also independent of its initial density profile within this reasonable range. Hence for our main suite of simulations, we can choose one of these as the cluster's initial density profile when trying to reproduce the mass and sizes of observed clusters in M83.

\begin{figure}
    \includegraphics[width=0.48\textwidth]{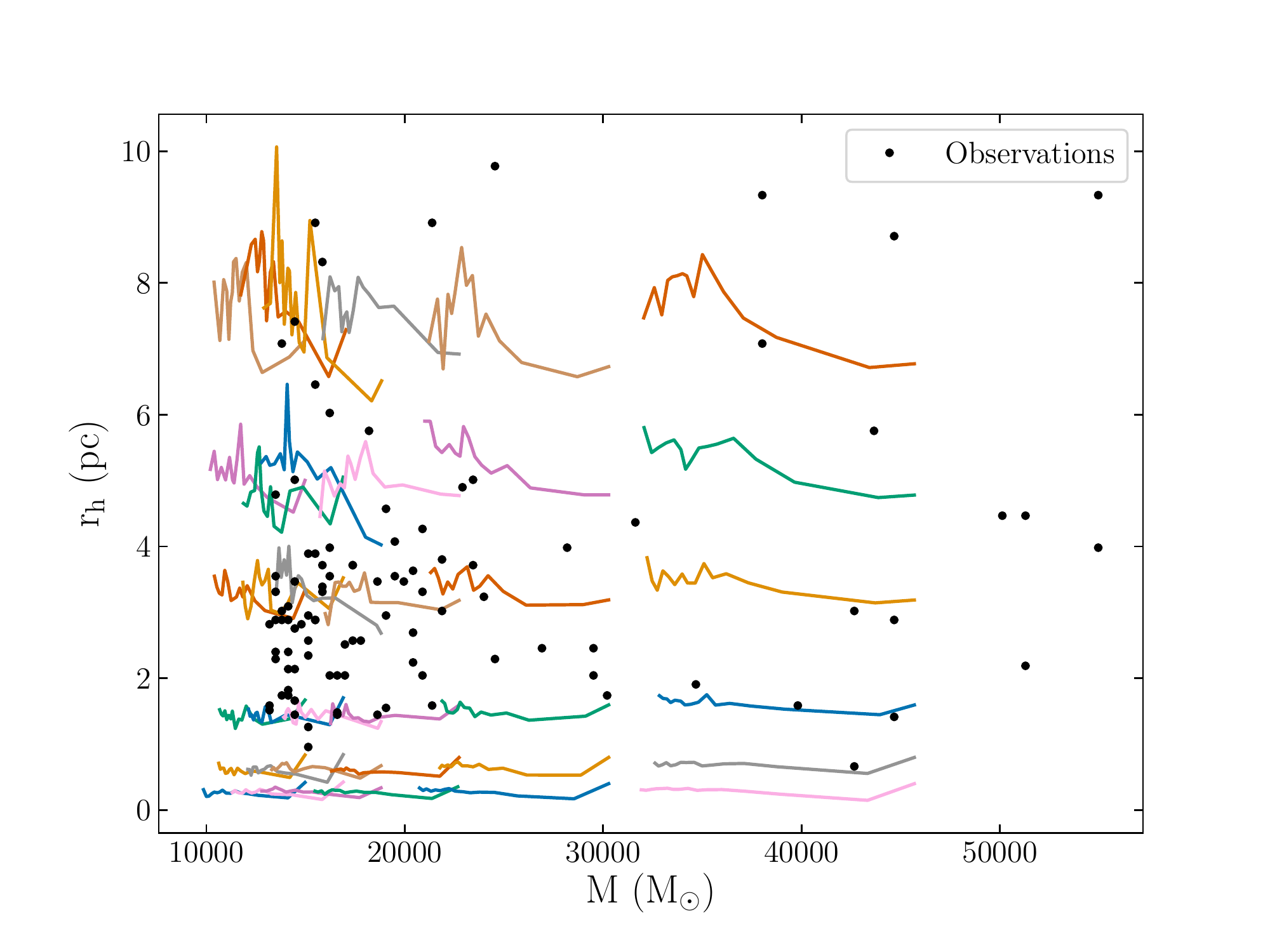}
    \caption{Mass and effective radius evolution of model clusters orbiting at 4 kpc with a range of initial masses and sizes. The observed M83 clusters have also been plotted (black points), illustrating that the models span the necessary range of the parameter space. %These models will be used for the ABC-MCMC fitting.
    }
    \label{fig:fullplot}
\end{figure}

Given the results of our tests on the effect that cluster orbit and initial density profile have on cluster evolution, it is only necessary to simulate clusters with a wide range of initial sizes and masses when working to determine the initial mass and size distribution of the M83 young massive cluster population. Hence we generate models of clusters orbiting at 4 kpc in the M83 galactic potential, with Plummer density profiles initially.
We ran models for clusters that initially consist of 25 000, 28 000, 31 000, 37 000, 50 000, and 75 000 stars and have initial half mass radii of either 0.5 pc, 1.0 pc, 2.0 pc, 4.0 pc, 6.0 pc, and 8.0 pc. The evolution of all models is illustrated in Figure \ref{fig:fullplot}, where we illustrate the evolution of the effective radius $r_h$ as a function of mass in order to best compare with the observed dataset. It is important to note that the $r_h$ evolution has been median smoothed over 10 Myr intervals since $r_h$ can fluctuate significantly between time steps due to the wide dynamical range encompassed by stellar luminosities. The observed clusters are also illustrated for comparison purposes.

There are a number of physical processes which could change the mass-radius evolution of N-body models that we have neglected in this work. A substantial population of primordial binary stars can help to heat globular clusters and prevent core collapse, but that process happens much later in the clusters' evolution. At these early times, the population of binary stars has quite a small effect on the evolution of the clusters' mass or radius  \citep{Chatterjee2013}.
Our models also have a very small black hole retention fraction. It is now understood that globular clusters may retain a population of a few hundred black holes, and they can form a dynamically important sub-system which can increase the size of the cluster, particularly its core. However, this process takes almost 1 Gyr in clusters of masses $\approx 5 \times 10^4$ M$_{\odot}$ even in the case of 100\% black hole retention \citep{Mackey2008} so it is not relevant at the young ages considered here.

\subsection{ABC-MCMC}\label{sec:abcmcmc}

The goal of this work is to take our suite of cluster models and determine how many of each mass/size combination are required to match the observed distribution of young clusters in M83. In other words, what distribution in the initial mass -- initial size plane will evolve into the present-day distribution? A brute-force approach would involve generating initial populations and then evolving all clusters forwards in time to see how well they reproduce the observed mass and size distributions, using for example a Markov Chain Monte Carlo (MCMC) method. Unfortunately this approach is difficult to do using direct $N$-body simulations as they are computationally expensive and could easily involve thousands of realizations of the M83 population. Fast cluster evolution codes do exist \citep[e.g.][]{Alexander2012}, but they are typically focused on the long-term evolution of clusters orbiting in galactic potentials that do not resemble that of M83. 

Instead, the problem of determining the initial distributions of mass and size of the clusters in M83 is well suited for the Approximate Bayesian Computation Markov Chain Monte Carlo (ABC-MCMC) algorithm \citep{Marjoram2003} (see also \citep{Gennaro2018} for a description of the algorithm). The general idea behind ABC-MCMC, in our context, is that we have an observed set of sizes $r_h$ and masses $M$ that we wish to reproduce using direct $N$-body simulations of star clusters from a set of initial conditions. To begin, we set up a population of star clusters from a randomly selected set of initial conditions. Specifically, we choose a slope for the initial cluster mass function; and assume a log-normal distribution of initial effective radii and choose a mean $\mu$ and dispersion $\sigma$ for that function. From those distributions, we select 101 pairs of initial mass and effective radii to create a single artificial cluster population. Guided by the suite of direct $N$-body simulations discussed in Section 2.2, we determine the $r_h$ and $M$ that each cluster will have after a random age drawn from a flat distribution up to 300 Myr (consistent with observed cluster ages in M83). If the simulated population is comparable to the observed population in both mass and size, to within a tolerance parameter, the parameters used to generate the initial population are accepted. Otherwise a new population is generated. 

We choose a power law initial cluster mass function with no truncation mass (i.e. not a Schechter function) for this work. There is disagreement in the literature whether a truncation mass is required, and typical values are larger than $10^5 M_{\odot}$ \citep[e.g.][]{Mok2019} which is above the maximum mass of our observed clusters. We are not sensitive to this high mass end, and so we remove this parameter from our fit. The initial mass function is sampled up to $1.1 \times 10^5 M_{\odot}$. We choose a log-normal distribution for the initial cluster effective radii, as we expect the initial distribution will likely have the same functional form as the final distribution in the absence of complete cluster disruption. 

To generate the first dataset to be used in the ABC-MCMC algorithm, we begin with $\alpha=-2$, $\mu=0.3$ and $\sigma=0.4$ for effective radii in pc. The dataset is then generated by selecting new values of $\alpha$, $\mu$, and $\sigma$ from Gaussian distributions with dispersions of $\sigma_{\alpha}=0.5$,  $\sigma_{\mu}=0.5$, and $\sigma_{\sigma}=0.2$. To match an initial mass - initial size pair to a given $N$-body simulation, we first find the cluster with the closest initial mass as the pair. The offset in initial mass is taken to be $\Delta M_i$. Of the $N$-body simulations with this initial mass, we then identify the simulation with the closest initial effective radius. The offset in initial size is taken to be $\Delta r_{h,i}$. Having already selected a random age for the initial mass - initial size pair, the mass and size of the cluster at its randomly selected age is then the mass and size of the matched $N$-body simulation with the offsets $\Delta M_i$ and $\Delta r_{h,i}$ added. If the final mass and size places the cluster outside of the range of masses and size of the observed clusters, it is ignored, and a new cluster is initialized. The process is repeated until 101 clusters have been generated. This final step of removing generated clusters that end up outside the observed mass-size range of M83 cluster results in the actual initial mass function of the generated clusters being slightly different than the initial mass function they were generated from. We therefore record the actual initial mass function of the generated cluster when estimating the initial mass function of the observed M83 clusters. The initial half-mass radii $r_{m,i}$ of the generated clusters are also determined by assuming their $r_{m,i}/r_{h,i}$ ratio is equal to the ratio of the model found in the cluster they were matched with. That ratio varies between 0.85 and 1.0.

Once a population of 101 clusters has been generated, we must evaluate how well it reproduces the observed dataset. For this step, following \citet{Gennaro2018} we first scale the masses and sizes of both datasets such that they range from 0 to 1. This scaling gives equal weight to the mass and size of the cluster. We then determine the kernel distance between the two datasets \citep{Phillips2011}, assuming a Gaussian kernel. As per \citet{Gennaro2018}, the kernel distance $\rho_K$ is calculated via:

\begin{equation}
\begin{split}
\rho_K^2(\mathcal{X},\mathcal{Y}) &= \sum_{\vec{x} \in \mathcal{X}} \sum_{\vec{x}^{\prime} \in \mathcal{X}} K(\vec{x},\vec{x}^{\prime}) + \sum_{\vec{y} \in \mathcal{Y}} \sum_{\vec{y}^{\prime} \in \mathcal{Y}} K(\vec{y},\vec{y}^{\prime}) \\
&- 2 \sum_{\vec{x} \in \mathcal{X}} \sum_{\vec{y} \in \mathcal{Y}} K(\vec{x},\vec{y})
\end{split}
\end{equation}

\noindent where $\mathcal{X}$ is the collection of observed cluster masses and sizes and  $\mathcal{Y}$ is the collection of model cluster masses and sizes. The function $K(\vec{x},\vec{y})$ is the kernel function, however since we are fitting over two parameters it is more accurately $K(\vec{x_1},\vec{x_2},\vec{y_1},\vec{y_2})$. We assume the kernel function to be the two-dimensional Gaussian kernel function, which is given by:

\begin{equation}
K(\vec{x_1},\vec{x_2},\vec{y_1},\vec{y_2})=\sum e^{-((\vec{x_1}-\vec{x_2})^2+(\vec{y_1}-\vec{y_2})^2)/\sigma^2}
\end{equation}

\noindent where we take $\sigma$ to be $3$ times the average minimum distance between the observed datapoints \citep{Gennaro2018}.

The general ABC-MCMC approach is to accept or reject a generated dataset based on whether or not the kernel distance is below a tolerance parameter. If accepted, the initial mass and size functions used to generate the dataset are used to reset the values of $\alpha$, $\mu$, and $\sigma$ and a new population is again generated from a Gaussian distributions with dispersions of $\sigma_{\alpha}$,  $\sigma_{\mu}$, and $\sigma_{\sigma}$. If rejected, $\alpha$, $\mu$, and $\sigma$ remained unchanged and new parameters are drawn from the initial Gaussian distribution. After a pre-set number of iterations, the initial mass and size functions that best reproduce the observed M83 clusters are taken to be the average of the $\alpha$, $\mu$, and $\sigma$ values accepted by the algorithm ($<\alpha>$, $<\mu>$, and $<\sigma>$).

We performed a convergence test to determine the tolerance parameter value that, below which, $<\alpha>$, $<\mu>$, and $<\sigma>$ no longer change. When performing 20,000 iterations, the values converged for a tolerance value of 200. Below 200 the fitting algorithm takes longer to run due to more cases being rejected without any change in $<\alpha>$, $<\mu>$, and $<\sigma>$. We also tested the method on artificial data with a known IMF and initial size function, and successfully recovered the expected values.

\section{Results and Discussion}

 We find that the initial cluster mass function for young clusters in M83 is well represented by a power-law with slope \alphafit\ $\pm$ \alphafiterror. As the cluster population evolves, the power-law slope of the simulated population's mass function becomes $-3.3 \pm 0.3$, consistent with stellar-evolution-only mass loss. Below 50 000 $M_{\odot}$ the evolved mass function has a slope of \alphafitevolve\ $\pm$ \alphafitevolveerror, identical to the observed population over the same mass range. The log-normal initial effective radius distribution that best reproduces the present day radius distribution has a mean $\mu$ of \rhlogfit\ $\pm$ \rhlogfiterror\ (corresponding to a radius of \rhfit\ pc $\pm$ \rhfiterror) with a dispersion $\sigma$ of \sigfit\ $\pm$ \sigfiterror. This initial effective radius distribution corresponds to an initial half-mass radius distribution with a mean of \rmfit\ and a dispersion of \rmfiterror. Once evolved, the clusters have a mean projected radius of \rhfitevolve\ pc, which is comparable to the observed value of 3.52 pc.

\begin{figure*}
    \includegraphics[width=0.9\textwidth]{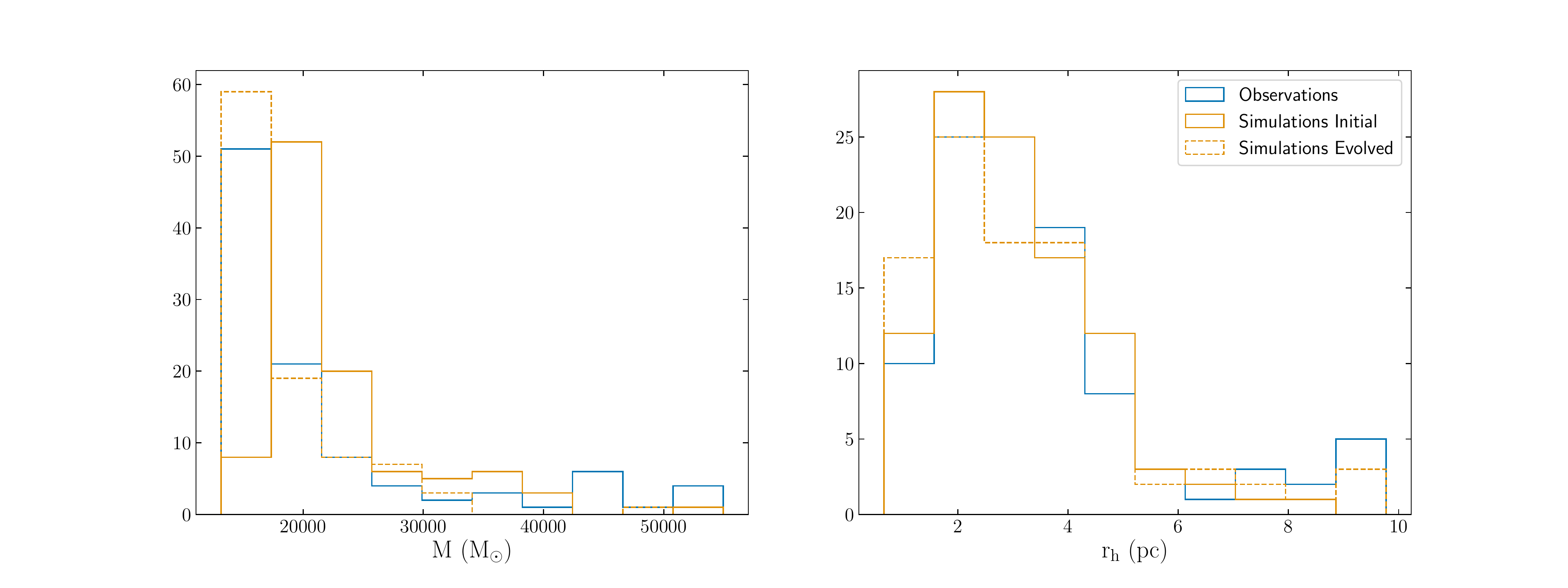}
    \caption{Present day distribution of observed and simulation cluster masses (left panel) and sizes (right panel). Observed clusters are illustrated in blue, while for simulated clusters we illustrate the initial (orange) and evolved (green) distributions. The mass function of a simulated population with an initial power-law slope of \alphafit will evolve into one with a slope of \alphafitevolve\ for $M < 50 \ 000 M_{\odot}$, precisely matching the observed mass function over the same mass range range. Similarly, with an initial size function that has a mean effective radius of \rhfit\ pc, simulated clusters evolve into a population with a mean effective radius of \rhfitevolve\ pc (comparable to the observed population with $<r_h> = 3.52 \ \rm pc$).}
    \label{fig:histplot}
\end{figure*}

To illustrate how these two distributions reproduce the observed distribution of clusters in M83, Figure \ref{fig:histplot} compares the observed mass and size of each cluster to a model population that has had their properties drawn from the best fit initial mass and size distributions and evolved for a random age. We draw 101 clusters from our best fit distributions and evolve them forward in time to create the `simulation' histograms. The left panel of Figure \ref{fig:histplot} illustrates that it is the low-mass end of the mass function that is responsible for minimizing the kernel distance between model and observed clusters, as the evolved simulated population closely resembles the observations below 50 000 $M_{\odot}$. As previously mentioned, the observed mass function is no longer well represented by a power-law above 50 000 $M_{\odot}$. Hence higher mass clusters are not reproduced as well given our base assumptions.

The right panel of Figure \ref{fig:histplot} illustrates that the best fit initial size distribution yields a present day distribution that is very similar to the observed cluster population of M83, with a mean $<r_h>$ that differs by only 0.26 pc. Smaller clusters are responsible for keeping the kernel distance down as there is a poorer agreement at the large $r_h$ end of the distribution. 

It is worth noting that the mass and size evolution of clusters in M83 appear to be independent of each other. We tested this by first fitting only the cluster radius distribution. The result is effectively the same as when mass is included as a parameter. Similarly if we fit only the masses, the best fit initial cluster mass function is a power-law of slope -2.8, only slightly steeper than when sizes and masses were fit together.

The lack of our results having any dependence on galactocentric distance is noteworthy, as \citet{Bastian2012} finds that when fitting clusters in M83 with a truncated mass function, inner and outer clusters have a different truncation mass. Similarly, \citet{Adamo15} finds different truncation masses and observed maximum cluster masses for M83 at 1, 2.5 and 4 kpc. However, we note that our observed clusters are mostly below these truncation masses, except possibly the proposed truncation masses at the largest galactocentric distances. Hence it appears that below the mass function's truncation mass (if it exists), neither a cluster's formation nor early evolution depend strongly on its location in the galaxy. 

The best fit log-normal initial size distribution has $\mu$ = \rhlogfit, which corresponds to a initial mean effective radius of \rhfit pc. This initial mean radius is smaller than the mean present day effective radii. The model clusters expand as a result of mass loss via stellar evolution, eventually reaching the same distribution as the observed population. A key underlying factor behind this comparison is our assumed age distribution. While we assume a flat age distribution between 3 and 260 Myr, if clusters are preferentially younger than the initial mean radius would be larger since clusters will not have as much time to expand. Similarly if clusters are preferentially older than they have started even more compact than \rhfit pc as they would have more time to expand. 

\begin{figure}
    \includegraphics[width=0.48\textwidth]{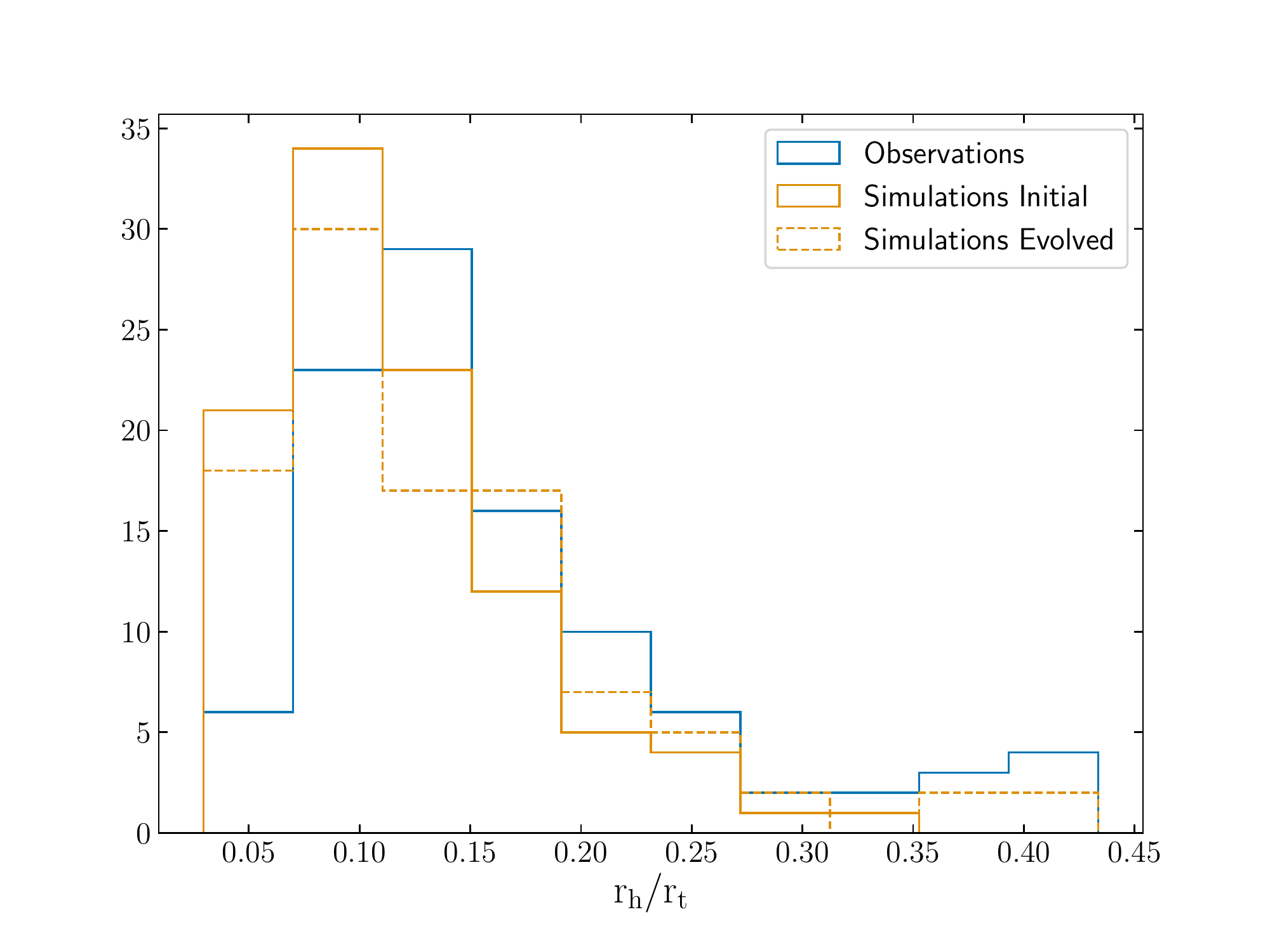}
    \caption{Present day distribution of observed and simulated cluster tidal filling factors. Observed clusters are illustrated in blue, while for simulated clusters we illustrate the initial (orange) and evolved (green) distributions).}
    \label{fig:rfplot}
\end{figure}

The relationship between a cluster's size and its local environment can be probed by calculating each cluster's tidal filling factor $r_h/r_t$, where $r_t$ is the tidal radius (also commonly referred to as the Jacobi radius). While various formalisms exist for the calculation of $r_t$, we make use of \citet{PortegiesZwart2010} in order to remain consistent with and compare our results to the observed M83 dataset \citep{Ryon2015}. In Figure \ref{fig:rfplot} we compare the distribution of $r_h/r_t$ from our best fit model dataset to the observed population.

Figure \ref{fig:rfplot} demonstrates that the evolution of a young massive cluster system generated from our best fit initial mass and size distributions will also have a similar $r_h/r_t$ distribution as the observed clusters. The observed population has a slightly larger average $r_h/r_t$ than the simulated population, with the simulation population also having a larger number of under-filling clusters with $r_h/r_t < 0.1$. These slight discrepancies are at first surprising, given the agreement between the individual mass and size distributions. However it is important to note that simulation clusters all orbit at 4 kpc and have a narrower $r_t$ distribution than the observed clusters which can have galactocentric distances as low as 0.5 kpc (and therefore very small $r_t$). Had we simultaneously tried to reproduce the mass-size-galactocentric radius distributions of clusters in M83, we would have found that smaller clusters are preferentially located closer to the centre of the galaxy. Since we find that early cluster evolution in M83 does not depend on galactocentric distance, the $r_h/r_t$ must therefore be imparted on the clusters during the cluster formation phase and may reflect the mass and size distribution of the progenitor giant molecular clouds.

\begin{figure}
    \includegraphics[width=0.48\textwidth]{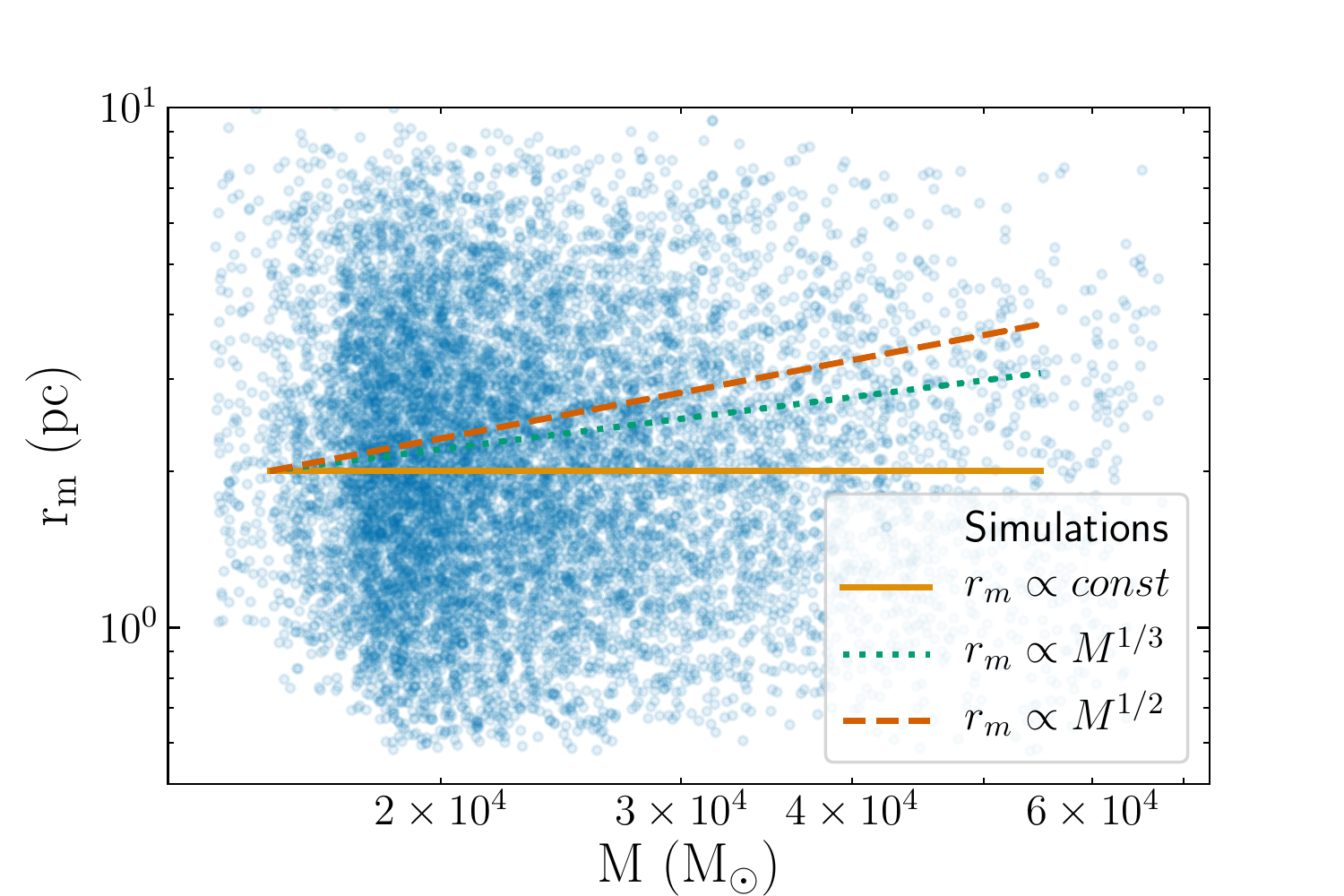}
    \caption{Initial mass-radius relation for 10 000 model clusters generated using our best fit initial mass function and initial half-mass radius function for young massive clusters in M83. For comparison purposes, lines of constant $R$ (orange), $R \propto M^{1/3}$ (green), and $R \propto M^{1/2}$ (red) are illustrated.}
    \label{fig:mrplot}
\end{figure}

The initial mass-radius relation for clusters is an important component of galaxy-scale cluster formation and evolution models. Some use a constant cluster radius for all masses \citep{Pfeffer2018}, while others allow the density to depend on mass for higher mass clusters, resulting in $r_m \propto M^{-1/3}$ \citep{Gnedin2014}, or assume a constant density for low mass clusters ($r_m \propto M^{1/3}$). This last exponent is more or less consistent with the observational compilation from \citet{Krumholz2019}, albeit with significant scatter. \citet{Choksi2019} provide a theoretical rationale for expecting the initial radius of clusters to depend on their mass according to $r_m \propto M^{1/2}$, and then demonstrate that an observed exponent of 1/3 can be found due to dynamical evolution of the low mass clusters, and a limit on high mass clusters which is dependent on the formation gas density. In this paper, we are explicitly calculating the dynamical evolution, and our clusters are too low mass for the environmental dependence to dominate. We look at the initial mass-radius relation from the best-fit initial mass and radius functions in Figure \ref{fig:mrplot} and overplot lines of constant radius with mass, $r_m \propto M^{1/3}$, and $r_m \propto M^{1/2}$. 

 There is significant scatter in Figure \ref{fig:mrplot}, with a large range in cluster radius as a function of mass. It appears that a single initial mass-radius relation is too simplistic. A formal fit to these data give a power-law slope of $\alpha=-0.02 \pm 0.02$, or consistent with no dependence of radius on mass. Repeating the analysis done in this paper on different galaxies and expanding to higher-mass clusters will help determine the correct mass-radius relation. Understanding the reason(s) for the scatter at constant mass will also be important. Given the young ages of our clusters, it is unlikely to be an effect of age and dynamical evolution. It is plausible that different galactic environments could allow for clusters of different sizes, depending on local tidal effects or the density of the local interstellar medium. More likely, the process of cluster formation itself might produce clusters with a range of sizes. Some simulations suggest that clusters are built up from mergers of smaller sub-clusters \citet{Howard2017}, and the number and properties of those mergers may result in different cluster structures. 

\section{Summary and Conclusions}

With the help of a large suite of $N$-body simulations, we make use of the ABC-MCMC method to determine the initial mass and size functions of the population of young star clusters in M83. We find that the cluster population can be produced by the evolution of star clusters that have a power-law initial mass function of slope \alphafit\ $\pm$ \alphafiterror and a log-normal initial effective radius distribution with $\mu =$ \rhlogfit\ $\pm$ \rhlogfiterror and $\sigma=$ \sigfit $\pm$ \sigfiterror in units of log(parsec). The corresponding initial half-mass radius distribution is also a log-normal distribution with $\mu =$ \rmlogfit\ $\pm$ \rmlogfiterror and $\sigma=$ \rmsiglogfit $\pm$ \rmsiglogfiterror in units of log(parsec).

We conclude that early dynamical evolution of star clusters is most important for their radius evolution, even over the first few hundred Myr, and that mass loss from stellar evolution drives an expansion of approximately 20\%. Since the first few hundred Myr of a cluster's life is also the time in which it is most likely to be influenced by environment through interactions with nearby molecular clouds, this expansion is an important process that should not be neglected in models of cluster formation and evolution. 

In these models, we have neglected gas expulsion, which likely occurs in the earliest stages of star cluster formation. Some authors \citep[e.g.][]{BanerjeeKroupa16} suggest that clusters form with sizes more like the widths of molecular gas filaments, $\approx$ 0.1-0.3 pc and then expand due to rapid gas explusion. More recent models of cluster formation \citep[e.g.][]{Howard2017, Lahen2020} suggest that clusters do not form monolithically, but instead build up over time through the combination of mergers of smaller clusters and the continued accretion of gas in molecular clouds. In these simulations, the cluster formation process ceases 5-10 Myr after the onset of star formation because gas is removed from the cluster vicinity through both accretion + star formation, and feedback. Our results from this paper suggest that whatever the process, clusters must have half-mass radii of approximately 2.5 pc very shortly after they become gas-free.

We also note that this value of 2.5 pc is only the mean of the log-normal distribution, and that clusters also have a range of initial sizes. Our dispersion of the best-fit log-normal distribution is \rmsigfit, which means the one-sigma range of sizes is $\approx$ 0.5 to 4 pc. This size range, and the fact that it seems not to be dependent on cluster mass, must also be explained by cluster formation models. And yet, our models suggest that the dynamical evolution of young clusters is not strongly affected by the galactic tidal field in M83, or the clusters' present-day location in the galaxy. It is probable that variations in the natal molecular clouds are responsible for this scatter, but those variations do not seem to be tightly tied to galaxy location. Detailed models of cluster formation in a range of formation environments are needed. 

\section*{Acknowledgements}
The authors would like to thank Jenna Ryon for providing the cluster data and further discussions. JW would like to further thank Jo Bovy for helpful discussions regarding ABC-MCMC and Enrico Vesperini for early feedback on the project. AS is supported by the Natural Sciences and Engineering Research Council of Canada. This work was made possible in part by the facilities of the Shared Hierarchical Academic Research Computing Network (SHARCNET: \url{www.sharcnet.ca}) and Compute/Calcul Canada.

\section*{Data Availability}
The observational data underlying this article was provided by \citet{Ryon2015}. The simulated data underlying this article will be shared on reasonable request to the corresponding author.

%%%%%%%%%%%%%%%%%%%%%%%%%%%%%%%%%%%%%%%%%%%%%%%%%%



\bibliographystyle{mnras}
\bibliography{M83} % if your bibtex file is called M83.bib

%%%%%%%%%%%%%%%%%%%%%%%%%%%%%%%%%%%%%%%%%%%%%%%%%%

% Don't change these lines
\bsp	% typesetting comment
\label{lastpage}
\end{document}